\documentclass[aps,prl,twocolumn,floatfix,citeautoscript,hyperref=pdftex]{revtex4-1}
\usepackage{graphicx}
\usepackage{dcolumn}
\usepackage{bm}
\usepackage{color}
\usepackage{amsmath}
\usepackage{amssymb}
\usepackage{multirow}
\newcommand {\beq} {\begin{equation}}
\newcommand {\eeq} {\end{equation}}
\newcommand {\bqa} {\begin{eqnarray}}
\newcommand {\eqa} {\end{eqnarray}}

\usepackage[colorlinks=true]{hyperref}

\begin{document}

\title{Perfect superconducting diode effect in altermagnets}

\author{Debmalya Chakraborty}
\thanks{Present address: Max Planck Institute for the Physics of Complex Systems, N\"othnitzer Stra\ss e 38, 01187, Dresden, Germany.}
\affiliation{Department of Physics and Astronomy, Uppsala University, Box 516, S-751 20 Uppsala, Sweden}

\author{Annica M. Black-Schaffer}
\affiliation{Department of Physics and Astronomy, Uppsala University, Box 516, S-751 20 Uppsala, Sweden}

\begin{abstract}

We investigate intrinsic superconducting diode effect in the unconventional superconducting state of $d$-wave altermagnets. We find large diode efficiencies in the wide-reaching finite-momentum pairing regimes of the phase diagram. Remarkably, even perfect diode efficiency of 100\% can be obtained in the presence of an external magnetic field. 
We attribute the largest efficiencies to competition between multiple zero-momentum (BCS) and finite-momentum superconducting states, which is connected to a topological nodal-to-nodeless transition in altermagnets with magnetic field, making our results applicable to a wide range of altermagnets. 

\end{abstract}

\maketitle

{\it{Introduction.}}--- Broken symmetries often enhance functionalities of materials. One example is broken inversion symmetry in semiconductors, resulting in the conventional diode effect \cite{Nagaosa24}. Recently it has been realized that a diode effect can also occur in superconductors with both broken time-reversal and inversion symmetries. In such a superconducting diode effect (SDE) a dissipation-less current flows in one direction, while only a normal current may flow in the reverse direction \cite{Nadeem23,Nagaosa24}. Due to its potential for constructing energy-efficient electronic devices, the SDE has received enormous attention, theoretically in both bulk superconductors \cite{Yuan22,Daido22,Ilicc22,Zinkl22,Scammell22,He22,Yuan21,Zhai22,Legg22,Takasan22,Karabassov22,Banerjee24a,Banerjee24,Hasan24,Roig24,bhowmik24,Chen24} and Josephson setups \cite{Hu07,Zhang22,Wei22,Pekerten22,Tanaka22,Kokkeler22,Davydova22,Baumgartner22,Cayao24,Debnath24,Cheng24}, and with many experimental findings \cite{Ando20,Narita22,Wu22,Bauriedl22,Lin22,Pal22,Trahms23,Ghosh24,Le24}, reaching a largest diode efficiency of 60\% in twisted flakes of cuprate $d$-wave superconductors \cite{Ghosh24}.

Early theoretical work \cite{Yuan22} already established the ubiquity of SDE in finite centre-of-mass momentum superconductors. In particular, finite-momentum superconductivity in the Fulde-Ferrell (FF) \cite{Fulde64} form, spontaneously breaks both time-reversal and inversion symmetries, hence generating SDE \cite{Yuan22}. Such a finite-momentum phase is found in non-magnetic materials in the presence of a magnetic field, but only in a narrow window of magnetic field and temperatures \cite{Fulde64,Larkin64}. Additionally, several correlated materials have experimentally been found to host finite-momentum superconductivity without applied magnetic field, then often called pair-density-waves (PDW) \cite{Hamidian16,Edkins19,Liu21,Kasahara20,Chen21,Gu23,Liu23,Zhao23}.
Finite-momentum superconductivity has also recently been theoretically discovered to be prevalent in altermagnetic materials for a  wide range of magnetic fields, even starting at zero field \cite{Rodrigo14, Shuntaro23, Chakraborty24,bose24, sim24, hong24} and also appearing as multiple phases interweaved with zero-momentum (or traditional BCS) superconductivity \cite{Chakraborty24}. In contrast to ferro- and antiferromagnetism, altermagnetism breaks the Kramer's spin-degeneracy but with a momentum dependent band spin splitting such that the net magnetization is  zero \cite{Smejkal20,Yuan20,Mazin21,Smejkal22a,Smejkal22,Hayami19,Hayami20} and has already been observed in many materials \cite{Feng22, Gonzalez23, Bai23, Olena24, Zhu24, Krempasky24, bai24,Jiang24,Zhang24a,Han24}.

Motivated by the prospects of finding finite-momentum pairing in altermagnets, we calculate in this Letter the SDE in altermagnets. We use a generic microscopic model of common $d$-wave altermagnetism, consider the simplest allowed superconducting states, and take into account the presence of an external magnetic field. We find large SDE efficiencies throughout the multiple finite-momentum pairing regions. Remarkably, we find diode efficiencies even reaching a perfect 100\%. We further uncover a direct connection between the largest diode efficiencies and a topological nodal-to-nodeless transition occurring in the altermagnet band spin splitting with applied magnetic field \cite{Fernandes24}. This connection to topology makes our results generic and applicable across a wide range of altermagnets. Our results provide a pathway combining unconventional magnetism and superconductivity for developing extremely energy-efficient electronic devices.

{\it{Model and method.}}--- We consider a minimal two-band model for a $d$-wave altermagnetic metal given by the Hamiltonian
\begin{eqnarray}
&&H_0=\sum_{k,\sigma} \left(\xi_{k}-\sigma (t_{\rm am}/2)(\cos(k_x)-\cos(k_y))+\sigma B \right) c_{k \sigma}^{\dagger} c_{k \sigma} \nonumber \\
&&+ \sum_{k,k',q} V_{k,k^{\prime}} c_{k+q \uparrow}^{\dagger} c_{-k+q \downarrow}^{\dagger}c_{-k'+q \downarrow}c_{k'+q \uparrow}, \label{eq:Hamil}
\end{eqnarray} 
where $c_{k \sigma}^{\dagger}$ ($c_{k \sigma}$) is the creation (annihilation) operator of an electron with spin $\sigma$ and momentum $k$, $\xi_{k}=-2t(\cos(k_x)+\cos(k_y))-\mu$ is the electron band dispersion, with $t=1$ as the energy unit, and $t_{\rm am}$ is the strength of the $d$-wave altermagnetic spin-splitting. We here only consider the two lowest energy bands in the prototype four-band tight-binding altermagnet model \cite{Smejkal22a}, since only they are relevant for superconductivity. These two bands encode a spin-sublattice locked dispersion, such that spin-up electrons only live on sublattice A, while spin-down electrons live on sublattice B \cite{Smejkal22a,Chakraborty24,Chakraborty24a}. Altermagnetism is generated from the combination of electric crystal field splitting and large spin-exchange field \cite{Smejkal22a}, where $t_{\rm am}$ in Eq.~\eqref{eq:Hamil}  mimics their combined effect. Further, $\mu$ is the chemical potential, which we tune to fix the average electron density $\rho=\sum_{k,\sigma}\langle c^{\dagger}_{k\sigma}c_{k\sigma}\rangle$. We also add an in-plane external magnetic field $B$ (with electron magnetic moment $\mu_B=1$) for controlling a Zeeman spin-splitting but with no orbital effects expected \footnote{The magnetic field does not affect $t_{\rm am}$ since its origin is different from a conventional magnet \cite{Chakraborty24}.}

Superconductivity is generated by the interaction $V_{k,k^{\prime}}$ in Eq.~\eqref{eq:Hamil}. Due to the spin-sublattice locking, uniform, or equivalently onsite, $s$-wave superconductivity is not present in a standard altermagnet \cite{Chakraborty24a}. We thus consider the simplest other pairing: spin-singlet pairing on nearest-neighbor bonds, giving either $d$-wave or extended $s$-wave superconductivity. Hence, we consider the generic effective attraction $V_{k,k^{\prime}}=-V\left(\gamma(k)\gamma(k')+\eta(k)\eta(k')\right)$, with $\gamma(k)=\cos(k_x)+\cos(k_y)$ and $\eta(k)=\cos(k_x)-\cos(k_y)$ being the two form factors for nearest-neighbor interaction on the square lattice and $V$ is a constant attraction strength. To proceed, we mean-field decouple the interaction in Eq.~\eqref{eq:Hamil} in the spin-singlet Cooper channel resulting in
\begin{eqnarray}
H_{\rm {MF}}&=&\sum_{k,\sigma} \xi_{k \sigma} c_{k \sigma}^{\dagger} c_{k \sigma} + \sum_{k} \left( \Delta^{Q}_{k} c_{-k+Q/2 \downarrow} c_{k+Q/2 \uparrow} + \textrm{H.c.} \right) \nonumber \\
&&+ \text{ constant},
\label{eq:Hamilmf}
\end{eqnarray}
where $\xi_{k \sigma}=\xi_{k}+\sigma (t_{\rm am}/2)(\cos(k_x)-\cos(k_y))+\sigma B$ and $\Delta^{Q}_{k}$ is the spin-singlet superconducting order parameter obtained by the self-consistency equation 
\begin{equation}
\Delta^{Q}_k=\sum_{k^{\prime}}V_{k,k^{\prime}} \langle c_{k^{\prime}+Q/2 \uparrow}^{\dagger} c_{-k^{\prime}+Q/2 \downarrow}^{\dagger} \rangle,
\label{eq:sc}
\end{equation}
with $Q$ being the finite center-of-mass momentum of the Cooper pair. We focus on both the $Q=0$ BCS superconductivity and FF phases with finite $Q$, where the phase of the superconducting order parameter varies but the amplitude does not. We can further use the ansatz $\Delta^{Q}_k=\Delta^{Q}_d\eta(k)+\Delta^{Q}_s\gamma(k)$, with $\Delta^{Q}_{d,s}$ being the $d$- or $s$-wave superconducting order parameters \cite{SudboBook}, both parametrically depending on $Q$ \cite{Chakraborty22b}. 

We proceed by solving the Hamiltonian $H_{\rm {MF}}$ in Eq.~\eqref{eq:Hamilmf} fully self-consistently using Eq.~\eqref{eq:sc} for fixed $Q$, and then obtain the true ground state by minimizing the ground state energy, $E=\sum_{k,\sigma}\xi_{k \sigma}\langle c^{\dagger}_{k\sigma}c_{k\sigma} \rangle-(\Delta^{Q}_d)^2/V-(\Delta^{Q}_s)^2/V+\mu\rho$ \cite{Waardh17,Chakraborty22b}, with respect to $Q$, resulting at an optimum $Q^*$. 
We then calculate the current using \cite{Cui06,altlandBook,MahanBook} $I=\sum_{k}\zeta_{k+Q/2}\langle c^{\dagger}_{k+Q/2 \uparrow}c_{k+Q/2 \uparrow}\rangle+\zeta_{-k+Q/2}\langle c^{\dagger}_{-k+Q/2 \downarrow}c_{-k+Q/2 \downarrow} \rangle$, where $\zeta_k=\sin(k_x)+\sin(k_y)$ and ignoring a universal prefactor. 
Importantly, our self-consistent minimization procedure ensures that the current in the ground state is always zero, due to Bloch's theorem \cite{Bohm49,Waardh17,Fulde64}. This result is also evident from the current being, by definition, proportional to the derivative of the free energy with respect to $Q$ \cite{Waardh17,Fulde64,Cui06}. 
However, an externally applied current produces a metastable state with Cooper pair momentum $Q$ around the optimum $Q^*$, i.e.~$Q=Q^*+\Delta Q$, with the sign of $\Delta Q$ set by the direction of current. We thus obtain the SDE by calculating the current for all $Q$ around $Q^*$ and use the thereby obtained maximum positive and negative currents to define the positive $I_{c}^{+}$ and negative $I_{c}^{-}$ critical currents, respectively \cite{Yuan22,Samokhin17}. It is here sufficient to consider only $Q>0$ due to the inversion symmetry present in the normal state \cite{Yuan22,Samokhin17} \footnote{It is possible that near $I_{c}^{-}$ there is a switch from the $Q=Q^*$ FF state to the $Q=-Q^*$ FF state \cite{Yuan22,Samokhin17}. However, such a branch switch occurs by forming inhomogeneous regions of the two FF states \cite{Samokhin17} separated by domain walls and depending on a system-dependent activation energy. Moreover, if such a branch switch occurs, transport becomes driven by dissipative domain wall motion \cite{Samokhin17} and hence $I_c^-$ is still the maximum dissipation-free current.}.
Using the common convention \cite{Nadeem23,Nagaosa24}, we finally define the SDE efficiency as $\eta=I_{c}^{+}-I_{c}^{-}/(I_{c}^{+}+I_{c}^{-})$ \cite{Yuan22} \footnote{We define $I_{c}^{-}$ as the absolute maximum value of the negative  current, in order to avoid using an absolute sign.}.

Below we report results for $V=2$ and $\rho=0.6$ at zero temperature and on a square lattice of size $1000\times 1000$, enough to mimic the thermodynamic limit and capture relevant values of $Q$. Since the ground state energy minima occurs for a uniaxial $Q$ \cite{Chakraborty24}, we only need results for $Q$ along the $x$-direction since the critical currents are obtained for momenta (anti)parallel to the FF momentum \cite{Yuan22}

\begin{figure}[t]
\includegraphics[width=1.0\linewidth]{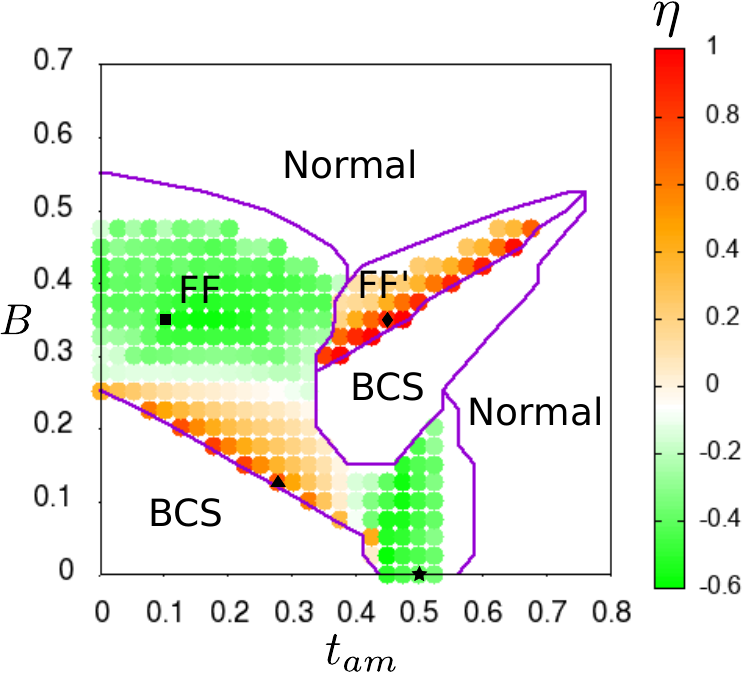} 
\caption{SDE efficiency $\eta$ in $B$-$t_{\rm am}$ phase diagram. Magenta lines demarcate different ground state phases, with the normal phase identified as $\Delta^{Q}_d<0.0009$ for all $Q$. BCS and normal phase have inversion symmetry, and hence no diode effect. Calculations are performed on a set of discrete points spaced $0.025$ apart in the $B$-$t_{\rm am}$ plane. Due to lack of numerical accuracy, $\eta$ is not shown in small white regions near the FF to normal transition. Black symbols ($\bigstar,\blacksquare,\blacktriangle,\blacklozenge$) marks parameters used in Fig.~\ref{fig:freeenergy}.
}
\label{fig:phasediagramdiode} 
\end{figure}

{\it{SDE in altermagnets.}}--- In Fig.~\ref{fig:phasediagramdiode} we show the superconducting phase diagram with computed SDE efficiency $\eta$ by varying both $B$ and $t_{\rm am}$. Before focusing on the SDE, we summarize the phase diagram, which almost takes the shape of a ``Yoda-ear" \cite{Chakraborty24}. The zero-momentum BCS state survives for low $t_{\rm am}$ and $B$, while finite-momentum FF superconductivity covers a large region of the phase diagram, including down to $B=0$ for a range of finite $t_{\rm am}$. Unexpectedly, the BCS state also reappears at higher $B$, as a field-induced BCS state, alongside with another finite-momentum state FF$^\prime$ with a different $Q^*$ from the FF state. 
Both FF$^{(\prime)}$ states in Fig.~\ref{fig:phasediagramdiode} spontaneously break time-reversal and inversion symmetries \cite{Waardh17,Fulde64,Cui06} and are thus expected to host SDE \cite{Yuan22}, as is also evident from the color-intensity plot of $\eta$ in Fig.~\ref{fig:phasediagramdiode}. In contrast, inversion symmetry is intact in the BCS state, such that no diode effect is present there (or in the normal state). 

\begin{figure*}[t]
\includegraphics[width=1.0\linewidth]{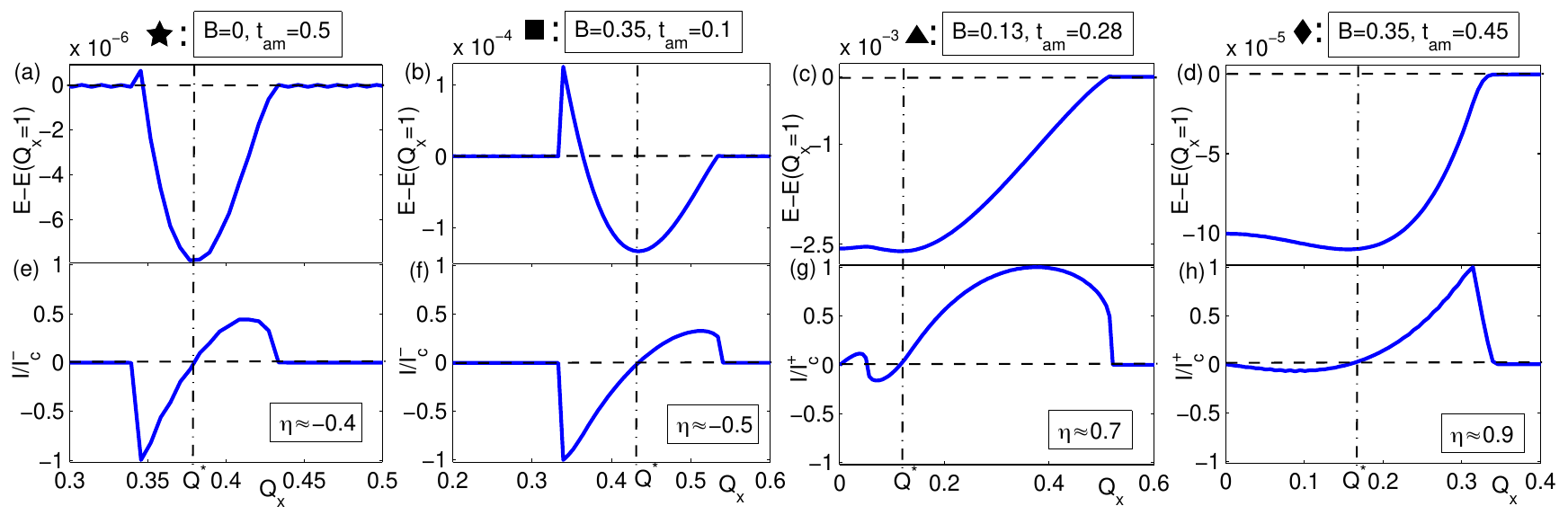} 
\caption{Energy (a-d) and current $I$ (e-h) for four representative points in the phase diagram Fig.~\ref{fig:phasediagramdiode}, indicated by black symbols ($\bigstar,\blacksquare,\blacktriangle,\blacklozenge$), as a function of $Q=(Q_x,0)$.  Energies given relative to the normal state by subtracting E($Q_x=1)$. $I$ normalized by max[$I_c^-,I_c^+$].
}
\label{fig:freeenergy} 
\end{figure*}

First focusing on the already well-known FF phase without altermagnetism, i.e.~$t_{\rm am}=0$, we find  $\eta>0$  for a small window of $B$ close to the BCS transition, but then $\eta<0$ for all higher $B$ in the FF state, but never reaching beyond $|\eta|=50\%$. 
Additionally, we also find $\eta \sim -50\%$ at zero applied field, $B=0$, for a range of altermagnetism strengths. 
However, the more interesting behavior appears when both $B\ne 0$ and $t_{\rm am}\ne 0$. Broadly, the diode behavior can be classified into two regions, characterized by the sign of $\eta$. Deep inside the FF state, we find $\eta<0$ indicating $I_{c}^{-}>I_{c}^{+}$ and a reverse diode effect. We here find a maximum negative efficiency of 60\%. However, in the FF phase near the BCS phase at both low $B$ and in the FF$^\prime$ phase at higher $B$, we find $\eta>0$, indicating a forward diode effect, and also significantly larger than in the reverse regime. 
Remarkably, we find the maximum $\eta$ reaching a perfect SDE efficiency of 100\% in the FF$^\prime$ phase near the transition to the field-induced BCS phase, but also the FF region at low $B$ can achieve a large $\eta \approx 75\%$. 
Taken together, both low and high applied magnetic fields in a superconducting altermagnet gives very large SDE efficiencies, including perfect SDE.

The large diode efficiencies, including perfect efficiency, can be understood by looking at the evolution of the energy $E$ as a function of $Q$. In Fig.~\ref{fig:freeenergy} we plot $E$ (relative to the normal state) with varying $Q=(Q_x,0)$ (a-d) and the calculated currents $I$ (e-h) for four representative points in the phase diagram, indicated by black symbols ($\bigstar,\blacksquare,\blacktriangle,\blacklozenge$) in Fig.~\ref{fig:phasediagramdiode}. We see directly that for all panels $I=0$ at the optimum $Q^*$ where $E$ reaches its global minima, fully consistent with $Q=Q^*$ corresponding to the ground state \cite{Waardh17,Fulde64,Cui06}. The critical currents, and hence $\eta$, is then obtained by finding the maximum current around this ground state. Thus the slope of $E$ as a function of $Q$ around $Q^*$ becomes an essential property.
In Figs.~\ref{fig:freeenergy}(a,b) the slope of $E$ for $Q_x>Q^*$ is less than the slope for $Q_x<Q^*$. Consequently, $I_{c}^{-}>I_{c}^{+}$ in Figs.~\ref{fig:freeenergy}(e,f), giving $\eta<0$. In contrast, in \ref{fig:freeenergy}(c,d) the slope of $E$ for $Q_x>Q^*$ is much larger than the slope for $Q_x<Q^*$. Hence, $I_{c}^{+} \gg I_{c}^{+}$ in Figs.~\ref{fig:freeenergy}(g,h), giving large positive $\eta$. 

In both Figs.~\ref{fig:freeenergy}(c,d), the FF$^{(\prime)}$ states are very close to a BCS state, reflected in the small energy difference between $E(Q=0)$ to $E(Q=Q^*)$ and also seen in Fig.~\ref{fig:phasediagramdiode}. However, the nature of large diode effect in these two regions are still quite different. 
In the FF phase in Figs.~\ref{fig:freeenergy}(c,g) both $E(Q=0)$ and $E(Q=Q^*)$ features minima, with $|E(Q=Q^*)|>|E(Q=0)|$ resulting into the FF ground state. The transition into the nearby BCS state occurs when $|E(Q=Q^*)|<|E(Q=0)|$ but notably both minima at $E(Q=0)$ and $E(Q=Q^*)$ remains also in the BCS state. As a result, $I_c^{-}$ never fully approaches zero, and hence we always find $\eta<75\%$, albeit this is already a substantial efficiency. In contrast, the transition of the FF$^\prime$ phase in Figs.~\ref{fig:freeenergy}(d,h) to its nearby field-induced BCS phase is different. As seen in Fig.~\ref{fig:freeenergy}(d), $E(Q=0)$ is a now maxima and thus the transition to the nearby BCS state occurs by this maxima becoming a minima, while, importantly, the minima at $E(Q=Q^*)$ no longer remains. Hence, $I_c^{-}$ asymptotically must reach zero near the FF$^\prime$ to BCS transition and thus $\eta$ reaches a perfect 100\%.

\begin{figure}[t]
\includegraphics[width=1.0\linewidth]{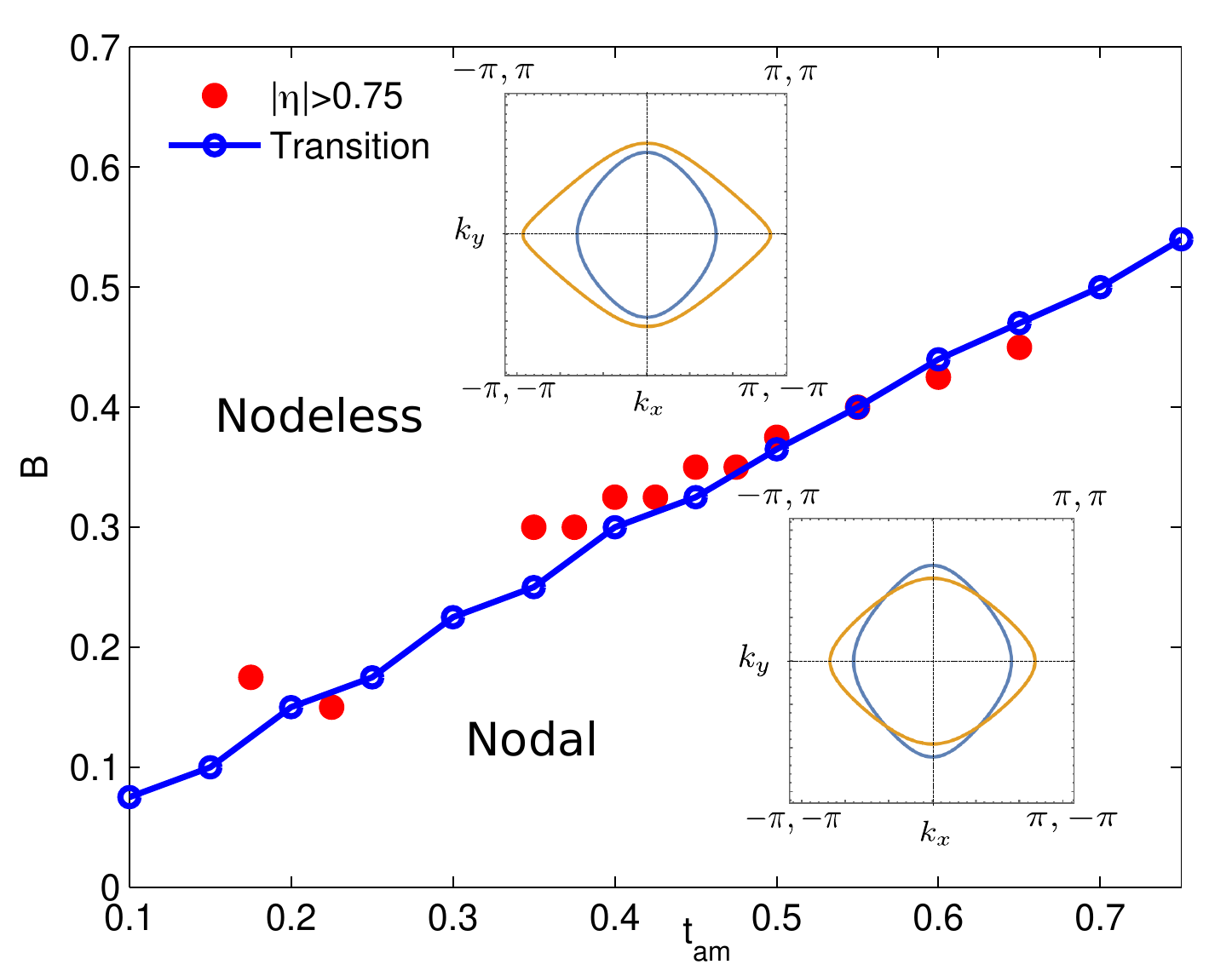} 
\caption{Critical $B$ (blue open circles, with line guide to the eye) for topological nodal-to-nodeless transition of the normal-state altermagnet band structure as a function of $t_{\rm am}$.  Red dots indicate regions of phase space with $|\eta|>75\%$. Critical $B$ is obtained for a fixed $\mu=-1.12$.
Insets: Contour plots of the normal-state electronic bands $\xi_{k \uparrow}=0$ (yellow) and $\xi_{k \downarrow}=0$ (blue) mapping the spin-split Fermi surfaces, with nodes for $B=0.1,t_{\rm am}=0.5$ (lower inset) and no nodes for $B=0.55,t_{\rm am}=0.5$ (upper inset).
 }
\label{fig:topnod} 
\end{figure}

{\it{Topological transition.}}--- The large, and even perfect, SDE in Fig.~\ref{fig:phasediagramdiode} may at first seem fortuitous, but we are able to directly connect it to a recently identified topological nodal-to-nodeless transition present in altermagnets when driven by an external magnetic field \cite{Fernandes24}.
In altermagnets, the Fermi surface gets spin-split with a momentum dependent splitting, see lower inset in Fig.~\ref{fig:topnod} for a $d$-wave altermagnet, with nodes along $|k_x|=|k_y|$ for $B=0$ where the Fermi surface is spin degenerate. With an applied magnetic field the nodes move towards $k_x=0$, due to the Zeeman effect. At a critical $B$, the nodes coalesce and for $B$ even larger there are no nodes (i.e.~nodeless) as the two spin Fermi surfaces do not intersect, see the upper inset Fig.~\ref{fig:topnod} where it is also clear that the spin Fermi surfaces now instead look more similar to a ferromagnet than an altermagnet. 
This nodal-to-nodeless transition is a topological transition, which has been shown to be independent of the considered band structure \cite{Fernandes24}. The main plot in Fig.~\ref{fig:topnod} show the magnetic fields where nodal-to-nodeless transition occurs (blue dots) in our system. 

Near the nodal-to-nodeless transition the two spin Fermi surfaces of the altermagnet become near-degenerate around $k_x=0$. This is also the region where a $d$-wave superconducting order parameter has its maximum values. As a result, the BCS state can most effectively compete with finite-momentum states in this transition region. We indeed find that $E(Q=0)$ in Fig.~\ref{fig:freeenergy}(d) can go from being a maximum to being a minima while in Fig.~\ref{fig:freeenergy}(c) we can instead achieve $|E(Q=Q^*)|<|E(Q=0)|$, thus entering BCS state in both cases with changing the magnetic field within this transition region. It is due to this presence of a nearby BCS state that $I_c^-$ is extremely small in the FF$^{(\prime)}$ phases. Hence we find large diode efficiencies, including the perfect efficiency, near the topological transition. 
We illustrate this in Fig.~\ref{fig:topnod} by plotting all areas  in the phase diagram with $|\eta|>75\%$ (red dots). The closeness of the red dots to the blue topological transition line clearly highlights the close connection of the large diode effect and the topological transition. We here also emphasize that large $\eta$ is not achieved in the FF phase near the BCS transition when $t_{\rm am}=0$. This is because this BCS to FF transition is a first-order transition, which is different from the ones described above.

{\it{Discussion.}}---In summary, we find a large intrinsic superconducting diode effect in the finite-momentum superconducting state of altermagnets, especially in the presence of external magnetic fields. The finite-momentum phase is classified into two broad regions, depending on sign of the diode efficiency. 
Already this finding can be used to experimentally detect different natures of the finite-momentum phase. 
We further show that the maximum diode efficiency in the traditional magnetic field-induced finite-momentum phase (i.e~in the absence of altermagnetism) is around 50\%, as is the maximum value achieved in altermagnets in the absence of magnetic field. In contrast, when both altermagnetism and magnetic field are present, we often find much larger diode efficiencies, including the remarkable possibility of reaching a perfect diode efficiency of 100\%. 

We identify the mechanism behind the large diode efficiencies by investigating the total energy with varying Cooper pair momentum. Our analysis reveals that a close competition between zero-momentum BCS and  finite-momentum superconducting ground states generates the largest diode efficiencies. We further connect this competition, and hence large diode efficiencies, to a topological nodal-to-nodeless transition present in altermagnets in an external magnetic field. The connection to a topological transition makes our results generic and not tied to specific altermagnet models or parameters. As an example, considering much weaker superconducting pairing strengths, the BCS region at low magnetic fields shrinks in size in the phase diagram, while the field-induced BCS state makes up a larger part \cite{Chakraborty24}. As a consequence, the region with perfect diode efficiency, which occurs near the topological transition of the finite-momentum pairing state to the field-induced BCS state, is enhanced. This example also illustrates that achieving large diode efficiency does not require using high magnetic fields. 

We further emphasis that our results are reached using a fully self-consistent treatment of superconductivity. Due to a strong spin-sublattice coupling in altermagnets, care must also be taken when considering possible superconducting pairing symmetries \cite{Chakraborty24a}. Self-consistently calculating  the chemical potential, assuming a fixed filling, is also required to reach the correct finite-momentum state.

{\it{Note added:}} When finalizing this manuscript, a preprint \cite{sim24} appeared investigating the SDE in an altermagnet in the absence of external magnetic field within a $s$-wave superconducting state using Ginzburg-Landau theory. While these zero-field results are similar to ours, our self-consistent results are valid for all magnetic fields and also deep within the superconducting phase.

\begin{acknowledgments}
{\it{Acknowledgments.}}--- We gratefully acknowledge financial support from the Knut and Alice Wallenberg Foundation through the Wallenberg Academy Fellows program KAW 2019.0309 and the Swedish Research Council (Vetenskapsr\aa det) grant agreement no.~2022-03963. The computations were enabled by resources provided by the National Academic Infrastructure for Supercomputing in Sweden (NAISS), partially funded by the Swedish Research Council through grant agreement no.~2022-06725.
\end{acknowledgments}

 \bibliographystyle{apsrev4-1}
\bibliography{Cuprates}

\end{document}